\documentstyle[aps,prl,multicol,graphics]{revtex}

\newcommand{\refref}[1]{Ref.\ \cite{#1}}
\newcommand{\figref}[1]{Fig.\ \ref{#1}}

\begin{document}
\title{Critical slowing down in polynomial time algorithms}
\author{A. Alan Middleton}
\address{Department of Physics, Syracuse University, Syracuse, NY 13244}
\date{\today}
\maketitle

\begin{abstract}
Combinatorial optimization algorithms which compute exact
ground state configurations in disordered magnets are seen to
exhibit critical slowing down at zero temperature phase transitions.
Using arguments based on the
physical picture of the model, including vanishing stiffness
on scales beyond the correlation
length and the ground state degeneracy,
the number of operations carried out by one such algorithm,
the push-relabel algorithm for the random field Ising model,
can be estimated. Some scaling can also be predicted
for the 2D spin glass.
\end{abstract}

\begin{multicols}{2}

There has long been a close link between the concepts of statistical
physics and the algorithms used to simulate condensed matter
systems. The fundamental connection is the mathematics of graphs, which
is applied analytically, for example, to
compute high temperature series.
Using sophisticated connectivity algorithms
borrowed from computer science, one can numerically calculate
quantities in percolation
to a high precision \cite{NewmanZiff}. 
The Fortuin-Kastelyn cluster representation
\cite{FortuinKastelyn} for magnets
is the basis for the Swendsen-Wang
algorithm \cite{SwendsenWangLiSokal}.
By implementing nonlocal dynamics, these algorithms can greatly
reduce the time needed for simulations near a phase transition.
The study of disordered
systems, such as spin glasses, pinned vortices in superconductors, and
random field magnets, lead to the introduction of
graphs with nonuniform edges \cite{Young,AlavaEtal}.
Early in the study of disordered
systems, it was realized that the study of such graphs is directly
related to issues of computational complexity.
In some cases \cite{Barahona}, computing the ground state
of a disordered material is computationally intractable, as the
relevant optimization questions on a graph are
NP-hard \cite{Papadimitriou}. Quite interestingly, some computationally
intractable problems have been found to have phase transitions
\cite{MonassonEtal}.  For
example, 
given an ensemble of logical expressions
characterized by the number of Boolean variables $N$
and clauses $M$, the fraction that are satisfiable
can exhibit finite size scaling
about a critical value of $M/N$.

In this letter, I present results on the behavior of ground state
algorithms near phase transitions in two models, the random field
Ising magnet (RFIM) and the 2D spin glass (2DSG), which, in contrast
with problems such as satisfiability,
are always solvable in time polynomial in the size of the graphs.
These phase transitions lead to singularities in the mean time
to solve the graph optimization problems.  For the RFIM,
a close connection is made between the critical slowing down of the
ground state algorithm, the correlation lengths, and the degeneracy of
the ground state in the thermodynamic limit.
Numerical evidence is presented that suggests
that the dynamic critical exponent is $z\approx 1$ for the RFIM;
scaling arguments suggest $z\ge 1$.
The behavior far from the transition can also be
explained.
The algorithm for the 2DSG is more difficult
to analyze, but the observed uniform time
per spin in the ferromagnetic (FM) phase is seen to be natural.

Generally, the dynamics at low temperature $T$ is
exceedingly slow in disordered magnets, as seen in
experiment and in Monte Carlo simulations using local moves \cite{Young}.
The glassy dynamical behavior is attributed to the complex
structure of the energy functional.
The free energy
barriers to equilibration at a length scale $\ell$ scale as $\ell ^
\psi$, so that the time to equilibrate a portion of the sample of
size $\ell^d$ are expected to
scale as $\sim \exp(\ell^\psi/T)$.  However, for some
random magnets there are algorithms which take time polynomial in $N$ to find
an {\em exact} ground state.  The process of finding the ground state
uses "nonphysical" configurations or moves: a local search or
simulated annealing that uses only physical configurations
and local moves is hindered by the large energy barriers.

When using local Monte Carlo moves at finite $T$ to model uniform magnets,
the relaxation or correlation time at continuous
transitions scales as $L^z$, with $z_{\rm loc} \ge \gamma/\nu$
\cite{Kawasaki}. Nonlocal cluster
moves, such as used in the Swendsen-Wang
algorithm, can reduce the
critical exponent $z$, with $z_{\rm cl} \ge\alpha/\nu$
\cite{SwendsenWangLiSokal}, with
$\alpha$ and $\nu$ the exponents for the heat capacity and correlation
length, respectively.
As the algorithms used to find
the ground states of disordered magnets do not use local moves and are not
{\em designed} to utilize clusters
related to a critical point, it is
less clear how many operations are required.
Polynomial bounds on the worst case behavior of these
algorithms do exist. For graphs with $n$ vertices and $m$ edges,
the highest level version of the push-relabel (PR) algorithm,
used here for the RFIM, will use $O(n^2\sqrt{m})$ time.
Useful algorithms for general matching \cite{CookRohe},
applied here to the 2DSG, 
use from $O(n^{3/2})$ to $O(n^3)$ operations, up to logarithmic
corrections, assuming $m\propto n$. In practice, however,
for typical disorder realizations of physical interest in finite
dimension $d$, these algorithms
are much faster, with
the average running time for many problems
scaling roughly as $N^{q}$, with $q$ typically in the range
$1.1-1.3$  \cite{AlavaEtal,MiddletonRBIMZengAAMShapir}.
For systems with a single $T=0$ fixed
point, such as the elastic medium
in a random potential \cite{MiddletonRBIMZengAAMShapir},
this scaling for the time would not be expected to vary with
parameters.

\begin{figure}
\begin{center}
\resizebox{0.8\linewidth}{!}
{\includegraphics{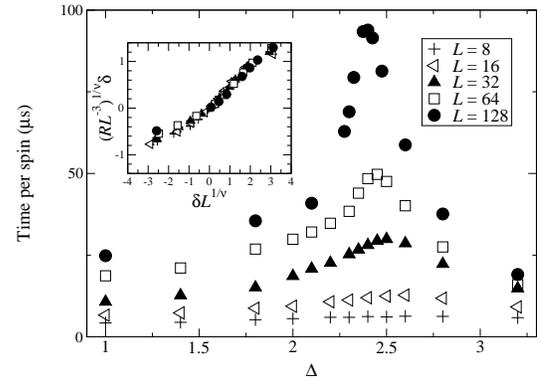}}
\end{center}
\caption{The CPU time needed to find the ground
state in the 3D RFIM, for a 766 MHz PIII.
The inset plots $(RL^{-3})^{1/\nu}\delta$, the scaled number
of relabel
operations per site, with $\delta = (\Delta - \Delta_c)/\Delta_c$,
vs.\ the finite-size scaling variable $\delta L^{1/\nu}$. The
values $\Delta_c=2.27$ and $\nu=1.37$ are not fit parameters, but
are derived independently. Statistical
error bars are about $1/5$ of the symbol size in both plots.}

\label{figtiming}
\end{figure}

The $T=0$ RFIM has a phase transition that has been extensively
studied\cite{OgielskiHartmannNowakFisherMiddleton} using a mapping
\cite{AngleEtal} of the ground state to the optimization problem max-flow
\cite{CormenEtal}.
The Gaussian RFIM has Hamiltonian
$H_R = -J \sum_{<ij>} s_i s_j - \sum_i h_i s_i$,
where the
allowed spin values on $N=L^d$ lattice sites $i$ are $s_i=\pm 1$, the
ferromagnetic couplings $J$ are positive, and the random fields $h_i$
are Gaussian distributed, with mean $0$ and variance $\Delta^2J^2$.
In the paramagnetic phase ($\Delta > \Delta_c \approx 2.27$),
the spins are pinned by the external fields $h_i$ and the magnitude of
the net magnetization $m=N^{-1}\sum s_i$ vanishes as $L\rightarrow 0$.
In the ferromagnetic phase ($\Delta < \Delta_c$), the ferromagnetic
coupling $J$ dominates and $|m|$ has a finite limit, with the
sign of $m$ determined roughly by the total of $\sum h_i$
(some spins will be reversed by fluctuations in $h_i$.)
Results for the
time for this algorithm to find the ground state in a 3D cubic lattice
with periodic boundaries
are plotted as a function of $\Delta$ and linear system size $L$ in
\figref{figtiming}. Plots of the number of primitive operations executed
are nearly identical in form. Near the phase transition separating the
ferromagnetic from paramagnetic phase, the time to find the ground
state shows a characteristic critical slowing down.
This notable
effect reflects the deep connection between the dynamics
of the ground-state algorithm and the physics of the RFIM.

The application of max-flow algorithms to the RFIM is well
established, but to make the connection between
scaling in the RFIM and algorithm timing, it is useful to review the
algorithm.  The network flow algorithm
generally used to solve the RFIM, because of its speed, is the
PR algorithm of Tarjan and Goldberg
\cite{TarjanGoldbergCherkassky}. The algorithm described here uses
a modification that removes the need for source and
sink nodes \cite{nosinknote},
reducing memory usage and also clarifying
the physical connection.

The modified PR algorithm starts by assigning an ``excess''
$x_i$ to each lattice site $i$, with $x_i = h_i$. Residual capacity
variables $r_{ij}$ between neighboring sites are initially set to
$J$. A height variable $u_i$
is then assigned to each node via a global update step. In this global
update, the value of $u_i$ at each site in the
set ${\cal T} =\left\{j|x_j<0\right\}$ of negative excess sites
is set to zero.
Sites with $x_i \ge 0$ have $u_i$ set to the
length of the shortest path, via edges with positive
capacity, from $i$ to ${\cal T}$.

The ground state is found by successively rearranging the excesses $x_i$,
via ``push'' operations, and updating the heights, via ``relabel''
operations. When no more pushes or relabels are possible, a final
global update determines the ground state: those sites which are
path connected by bonds with $r_{ij}>0$ 
to ${\cal T}$ have $s_i=-1$, while the sites which
are disconnected from ${\cal T}$ have $s_i = 1$.  A push
operation moves excess from
a site $i$ to a
lower height neighbor $j$, if possible, that is,
whenever $x_i>0$, $r_{ij} > 0$
and $u_j = u_i-1$. In a push, the working variables are modified
according to $x_i \rightarrow x_i -
\delta$, $x_j \rightarrow x_j + \delta$, $r_{ij} \rightarrow r_{ij} - \delta$,
and $r_{ji} \rightarrow r_{ji} + \delta$,
with $\delta = \min(x_i, r_{ij})$.  Push operations tend to move the
positive excess towards sites in ${\cal T}$.  When $x_i > 0$ and no
push is possible, the site is relabeled, with $u_i$ increased to
$1 + \max_{\{j| r_{ij} > 0\}} u_j$.  In addition, if a set of highest
sites ${\cal U}$ become isolated, with $u_i > u_j+1$, for all $i\in{\cal U}$
and all
$j\notin{\cal U}$, the height $u_i$ for all $i\in{\cal U}$ is
increased to its maximum value, $N$, as these sites will always be
isolated from the negative excess nodes.  A proof of the correctness
of the PR flow algorithm can be found in standard textbooks
\cite{CormenEtal} and its application to the RFIM is
well known \cite{AlavaEtal}.
Periodic global updates, here applied every $N$ relabels,
are often crucial to the
practical speed of the algorithm.
The highest site heuristic is used here,
which applies pushes and relabels where $u_i$ is maximal and
$x_i > 0$.

The PR
algorithm is intuitively appealing: when the
initial capacities within a region are large, the excess can be
rearranged so that the positive excesses
cancel the negative excesses as much as
possible. The remaining
excess values having the sign of the original total excess for
the region. As $r_{ij} = J$ and $x_i = h_i$ initially,
large capacities correspond to the ferromagnetic bonds being strong
enough to favor alignments of the spins, with the spin direction given by
the sign of the
total $h_i$ for the region. If the initial capacities are not
large enough (weaker $J$), the regions align independently, according to the
local field, and the excesses do not cancel.
The number of steps needed to move excess across the
diameter of a region via push operations is bounded below by the
linear size of the region.  Note that the
running time in the FM phase is somewhat dependent
on the majority magnetization,
due to the up/down asymmetry of the algorithm.

I now argue that these correspondences can be used to bound
the running time of the algorithm, using the physical properties of
the RFIM, particularly the behavior of the ground state degeneracy in
the thermodynamic limit.  Recent
work \cite{OgielskiHartmannNowakFisherMiddleton} has given
strong numerical evidence of insensitivity of the interior spins in
the ground state to boundary conditions, when $\Delta > \Delta_c$.
(This is in contrast with the
scenario for, e.g., mean field \cite{Young} or highly disordered
\cite{NewmanStein8D} spin glasses in $d>8$, where the entire solution
is sensitive to the volume.)
Most importantly, this implies that
the ground state solution is determined by the $h_i$ within a volume
typically of the size of the correlation volume $\xi^d$. However, in
the FM phase, the interior {\em is} sensitive to the boundaries and
a finite fraction of the  spins flip infinitely often as the sample
volume is increased, as the net magnetization is given by a coarse-grained
global sum of the $h_i$.

First, consider the case $\Delta \gg 2d$, where, in the ground state,
almost all spins satisfy $s_i = {\rm sgn}(h_i)$.
In the algorithm, the initial positive excesses at
sites $i$ where $h_i>2dJ$ remain positive, as the total capacity of
bonds leaving the site is only $2dJ$, so that push operations can only
reduce $h_i$ by $2dJ$. The sites with excess $x_i < -2dJ$ also always
have negative excess. There will be some small rearrangement, but only
locally, and the sign of the excess will change only at a very few sites
during the execution of the algorithm, which terminates after a number
of operations $\propto N$ (up to logarithmic corrections
\cite{MiddletonUnpub}.)

A similar scenario holds, but at scales $\xi$, for $\Delta>\Delta_c$.
The algorithm establishes the boundaries of correlation volumes
by rearrangement of excess over distances of scale
$\xi$. Further rearrangement is blocked by the effective decrease in
residual capacity with scale (as the stiffness, corresponding to
the scale dependent $J$, decreases rapidly on scales
greater than $\xi$ \cite{percnote}.)  The question to be
answered, then, is how long the algorithm takes, per spin, to
redistribute excess on scale $\xi$.  The number of push and relabel
operations in a volume $\xi^d$ is bounded below by $\xi^{d+1}$.
For the
excess to be pushed over a distance $\xi$, the relative heights must
differ by at least $\xi$, so that at least $O(\xi^{d+1})$ relabels $R$
must be performed for each correlation volume
(global updates lead to height changes, but 
these do not appear to affect the scaling, empirically.)
This gives the estimate $R\sim L^3 \xi$.
This scaling is consistent with numerical results, up to logarithmic
corrections, for $d=1,3$. The inset to \figref{figtiming} show a scaling
plot for $R$, for example, with $(RL^{-3})^{1/\nu}\delta$ plotted as
a function of the finite size scaling variable $\delta L^{1/\nu}$,
$\delta = (\Delta-\Delta_c)/\Delta_c$,
with the values $\nu=1.37$ and $\Delta_c=2.27$ {\em fixed parameters,
determined independently} \cite{OgielskiHartmannNowakFisherMiddleton}.
A fit in $d=1$, with $\nu=2$ and $\Delta_c=0$ also
fixed by known results, describes the data
well for $L \le 5 \times 10^6$, with $R \sim L \xi \ln(L/\xi)$ (without
the global relabeling heuristic.)

\begin{figure}
\begin{center}
\resizebox{0.8\linewidth}{!}
{\includegraphics{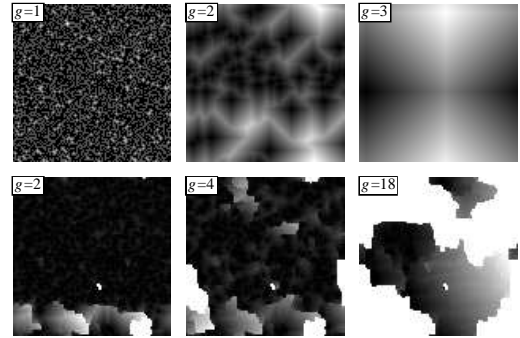}}
\end{center}
\caption{Images of the heights $u_i$
at intermediate stages of the PR algorithm,
in the 2D RFIM for $100^2$ samples,
for $\Delta=10^{-5}$
(top) and $\Delta=1.0$ (bottom). In the latter case,
coarsening is cutoff by the finite correlation
length; the visible spin-up domain shows some structure.
The shading is darkest at the maximum $u_i$ and is white
where $u_i=0$.
The number of global updates, $g$, is given by the labels.}
\label{figcoarsen}
\end{figure}

One limit where the asymptotic time for the algorithm can be
described more precisely is 
when $\Delta\ll[L\ln(L)]^{-1/2}$.) Here, the capacities do not
limit the rearrangement of excess: the final state is
either $x_i \ge 0$ everywhere
or $x_i \le 0$ everywhere, corresponding to a uniform $s_i=\pm 1$
state, according to whether $\sum h_i \ge 0$ or $\sum h_i \le 0$,
respectively. The dynamics of the PR algorithm is
set by the fluctuations in the total $h_i$ in regions at each
length scale. At any time scale of the computation, the positive
excess will be pushed towards the nearest negative excess region,
with the distribution of excess
negative or positive over a particular volume with $\ell^d$ sites, as the
rearrangement of excess will have been completed over shorter lengths at
an earlier time scale.
Generally, but especially
given sufficiently frequent global updates, the sites with greatest
$u_i$ will be furthest from the set $\cal T$. As the highest sites
are examined for pushing, the excess will be moved
from these sites to the next lower level. These sites will then have
their excess moved to the next lower level and the algorithm will
``sweep'' the excess through the volume of diameter $\approx 2\ell$. This will
establish a distribution of excess that will have uniform sign
over a region of size $2\ell$. In this fashion, the algorithm will
solve for the sum of $h_i$ recursively. The number of
steps at each stage will be $L^d$ and $O(\ln L)$ stages will
be required, giving a total running time $\sim L^d\ln(L)$.
This result is consistent with the numerical timings for very small
$\Delta$, with the
data for $RL^{-d}$ linear in $\ln(L)$
over sample dimensions $16 \le L \le 128$ to within the
$1\%$ numerical error for $d=3$ and over $10^3 \le L \le 5 \times 10^6$
to within the same error for $d=1$. 
Coarsening of the height variables during the algorithm is displayed
in \figref{figcoarsen}.

\begin{figure}
\begin{center}
\resizebox{0.8\linewidth}{!}
{\includegraphics{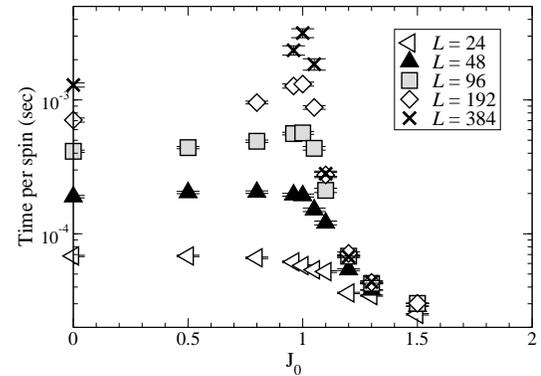}}
\end{center}
\caption{
The CPU time needed to find the ground
state in the 2D spin glass, as a function of the ferromagnetic
strength $J_0$ and system size $L$.}
\label{plot2dsg}
\end{figure}

To indicate the generality of critical slowing down and partial
arguments for other systems, it is useful to compare the RFIM results with
results for the 2DSG.
The Hamiltonian is
$H_S = \sum_{<ij>} J_{ij} s_i s_j$,
where again the $s_i$ are Ising spins.  The
Gaussian distributed $J_{ij}$ have mean $J_0$ and
variance $1$.  The SG to
FM transition \cite{McMillanSGFM} takes place
as $J_0$ is increased through the critical value $J_c \approx 0.96$.
The mapping
from the 2DSG with free boundaries
to a general matching problem is given in \refref{Barahona}:
energy is minimized by selecting a state with
a minimum total weight for frustrated bonds.
Timing results for the 2DSG are shown in \figref{plot2dsg}.
One notable difference from the RFIM is the apparent convergence to
{\em constant time per spin} in the FM phase. As
the algorithm used for the 2DSG uses a bond representation,
the algorithm does not need
to distinguish degenerate states related by global spin flips.
If the frustration is low enough, the FM phase is obtained by {\em local}
operations giving a solution with percolating
unfrustrated bonds.
For low $J_0$, in the SG phase, in contrast, though locally the
ground state is insensitive to the boundaries, the global ground
state is sensitive to the disorder and the operations (augmenting
paths) must therefore be carried out over all scales, so that the
time per spin may diverge as a power of $L$ (numerically, approximately
as $L^{0.78\pm0.10}$ for $96\le L \le 720$ and $J_0=0$.)
At the critical point, independent calculations \cite{MiddletonUnpub}
show that the domain wall fractal dimension increases slightly, so
that bond updates on large scales are more expensive.

In summary, the time for a polynomial time algorithm to find ground
states is examined near zero temperature critical points for two
models, the random field Ising model and the 2D spin glass.
At the critical points, the combinatorial optimization
algorithms used, PR flow and general matching,
while exponentially faster than, say,
simulated annealing at finding the ground state, slows down. This
slowing down of the algorithm is argued to be closely related to
important physical ideas, namely, the uniqueness or two-fold
degeneracy of the ground state in various phases
and the divergence of the correlation length as
the transition is approached. The time for the algorithm can be
understood in detail for the RFIM with small random fields.
It will
certainly be of interest to consider such slowing down and
scaling at other fixed points for other
polynomial time algorithms, to expand our
understanding of the physics and algorithms for these systems. These
considerations are clearly relevant for developing
efficient parallel algorithms.
I would like to thank J. Machta for useful discussions.
This work was supported
in part by the NSF (DMR-9702242.)

\end{multicols}


\begin{references}
\bibitem{nosinknote}
The use of negative excesses is also useful in min-cost flow algorithms.
\end{references}
\end{document}